A comparative analysis of transcription factor expression during metazoan embryonic development

Authors: A. N. Schep (1 and 2), B. Adryan (1)

( (1) Cambridge Systems Biology Centre, University of Cambridge, Cambridge, United Kingdom. (2) Department of Genetics, Stanford University, Stanford, California, United States)

Abstract

During embryonic development, a complex organism is formed from a single starting cell. These processes of growth and differentiation are driven by large transcriptional changes, which are following the expression and activity of transcription factors (TFs). This study sought to compare TF expression during embryonic development in a diverse group of metazoan animals: representatives of vertebrates (*Danio rerio*, *Xenopus tropicalis*), a chordate (*Ciona intestinalis*) and invertebrate phyla such as insects (*Drosophila melanogaster*, *Anopheles gambiae*) and nematodes (*Caenorhabditis elegans*) were sampled, The different species showed overall very similar TF expression patterns, with TF expression increasing during the initial stages of development. C2H2 zinc finger TFs were over-represented and Homeobox TFs were under-represented in the early stages in all species. We further clustered TFs for each species based on their quantitative temporal expression profiles. This showed very similar TF expression trends in development in vertebrate and insect species. However, analysis of the expression of orthologous pairs between more closely related species showed that expression of most individual TFs is not conserved, following the general model of duplication and diversification. The degree of similarity between TF expression between *Xenopus tropicalis* and *Danio rerio* followed the hourglass model, with the greatest similarity occuring during the early tailbud stage in *Xenopus tropicalis* and the late segmentation stage in *Danio rerio*. However, for *Drosophila melanogaster* and *Anopheles gambiae* there were two periods of high TF transcriptome similarity, one during the Arthropod phylotypic stage at 8-10 hours into *Drosophila* development and the other later at 16-18 hours into *Drosophila* development.

INTRODUCTION

During metazoan embryonic development, a single cell grows, divides, and differentiates into a complex organism with numerous distinct tissues. While the specifics of embryogenesis differ between animal species, all bilateria share certain features of development. Repeated cleavage divisions follow egg fertilization and activation. During gastrulation, the three major germ layers are formed, and during organogenesis those germ layers differentiate into the organs of the animal [1,2]. This process of embryonic development is driven by large transcriptional changes. *In situ* hybridization, microarrays, and more recently RNA-seq have been used to monitor gene expression over the course of development in a variety of model organisms and some non-model organisms as well [3–18]. These studies have shown that embryonic development is characterized by expression changes of most transcripts [4,11–13,19,20]. Furthermore, expression changes do not occur gradually over the course of embryonic development; rather embryonic development is characterized by periods of dramatic transcriptional changes and periods of relative expression stability [4,11–13,21]. Groups of genes can be clustered together based on when they change expression [4,13–15].

Many comparative transcriptomic studies have focused on the question of when expression of orthologous genes is most similar between different species. These studies have provided support for the "hourglass" model of development, which, since Haeckel's biogenetic law, posits that organisms are very different at the onset of embryogenesis but become more similar through a stage later in embryogenesis after which they once again diverge [22–25]. The stage at which the organisms are most similar is termed the "phylotypic" stage. The idea of a phylotypic stage was first based on the time period during development when organisms were most similar morphologically [23–25]. The hourglass model and the idea of a morphological phylotypic stage has been controversial [26,27], but transcriptomic evidence is now showing that the phylotypic stage is also when organisms are most similar in terms of gene expression [17,18,21,28].

Understanding the patterns of expression of transcription factors (TFs) in particular is important as the expression and binding of these regulatory proteins is crucial for controlling the transcriptional changes that occur throughout development. Levine and Davidson (2005) have proposed that the greater number of transcription factor genes relative to total genes in more complex organisms may enable their greater complexity [29]. A few studies have looked in depth at the expression profiles of large numbers of TFs during development. Imai *et al*. (2004) found that in *Ciona intestinalis* 74% of transcription factor genes were deposited maternally and 56% expressed zygotically, with zygotic expression of transcription factor increasing throughout embryogenesis. Forkhead (FOX) transcription factor genes tended to be expressed early in embryogenesis, while bHLH, Homeobox, and Ets family transcription factors tended to start being expressed later in embryogenesis [16]. Similarly, Adryan and Teichmann (2010) found that in *Drosophila melanogaster* 95% of transcription factors were expressed at some point during embryogenesis, with about 60% of transcription factors being maternally contributed. In contrast to *Ciona*, however, transcription factor expression did not increase throughout development but rather peaked 10-12 hours into development and then decreased over the last 12 hours of embryogenesis. Transcription factor expression was grouped into four broad classes: early expression only, late expression only, not maternal but continuous expression, and maternal and continuous expression. Analysis of transcription factor family expression showed that zinc finger transcription factors were strongly represented early in development while HLH, Homeobox, bZIP, forkhead, Ets, T-box, and GATA family transcription factor expression started low and increased over the course of embryogenesis [30].

This study sought to compare the patterns of transcription factor expression and transcription factor family utilization in representatives of vertebrates (*Danio rerio*, *Xenopus tropicalis*), a chordate (*Ciona intestinalis*) and invertebrate phyla (*Drosophila melanogaster*, *Anopheles gambiae, Caenorhabditis elegans*). The patterns of transcription factor expression and transcription factor family utilization in the different species were compared to determine whether these patterns were largely similar or if there were any differences between the species that correlated with differences in development and cell fate specification. Such comparisons are limited by different availabilities of data for different species and the difficulty of comparing expression measures from different technologies. While an online resource 4DXpress exists, which is aimed to facilitate comparisons in expression between species by bringing together *in situ* hybridization data from different species and enabling users to search for expression patterns of particular genes in different species [31], it is limited by the choice of species and the inherent difficulties in comparing spatio-temporal expression patterns between species with different anatomies and developmental timeframes. This study sought to bring together the best available expression data for embryonic development in a diverse set of species.

RESULTS & DISCUSSION

Similar trends of transcription factor expression throughout embryogenesis

Lists of transcription factors and structural domain assignments for each transcription factor for *Danio rerio*, *Xenopus tropicalis, Ciona intestinalis, Drosophila melanogaster, Anopheles gambiae* and *Caenorhabditis elegans* (Supplementary Figure S1) and were obtained from the DBD transcription factor prediction database [32]. In summary, 1000 TFs were analyzed for each of the vertebrate species, and about 500-600 TFs in case of the invertebrate species. Supplementary Figure S2 shows the number of transcription factors in each species that are classified as having Homeobox, C2H2 zinc finger, C4 zinc finger, Ets, bZIP, or HLH domains in each species considered. These transcription factor families were the focus of this study as they are the families present in large enough numbers in each organism to be able to make significant comparisons. *In situ* hybridization, single-color microarray, and RNA-seq data were used to determine the percentage of transcription factors expressed throughout embryogenesis in the species of interest (Figures 1 and 2).

In the zebrafish *Danio*, the *in situ* (Figure 1a) and microarray (Figure 2a) data show TF expression increasing throughout most of embryogenesis until the early pharyngula or late segmentation stages before declining slightly. The microarray data shows a much higher percentage of TFs being expressed initially than the *in situ* data (62% compared to 27%), but a similar proportion being expressed during the peak period of TF expression (approximately 90%). For *Xenopus tropicalis*, the trends in the microarray data (Figure 2b) are very similar to those observed from the microarray data for zebrafish. TF expression increases throughout most of development from 65% of TFs being expressed in the fertilized egg to 93% of TFs being expressed during the middle of the tailbud stage, with TF expression then decreasing slightly during the late tailbud stage. In *Ciona*, the trends from the *in situ* hybridization data from the Ghost database are mostly similar except that TF expression is initially very high (70%) but then decreases dramatically between the fertilized egg and 16-cell stages to 23% before gradually increasing throughout the remainder of embryonic development, reaching 62% at the middle-late tailbud stage (Figure 1b).

The findings of Adryan and Teichmann (2010) with respect to the percent of *Drosophila melanogaster* TFs expressed over time were reproduced for this study, and the novel RNA-seq data from modENCODE [3,4,19] confirms their original results based on RNA *in situ* hybridization and microarray. All three data sets show TF expression increasing during the first half of embryonic development. The RNA-seq data, which extends throughout the first 24 hours of development, shows that TF expression declines during the last third of embryogenesis. A very similar pattern was observed for *Anopheles gambiae* based on microarray data. The percentage of TFs expressed increases from 50% to about 65% throughout the first half of embryogenesis before decreasing to only about 40% of TFs being expressed. This decrease in TF expression occurred earlier and was more drastic than any decreases in TF expression observed in the other, non-insect species; this decrease in TF expression midway through embryogenesis could therefore represent an insect specific feature that precedes hatching into the larval stage.

In *Caenorhabditis elegans*, microarray data from Levin et al. (2011) indicate that about 40% of TFs are expressed at the 4-cell stage and that TF expression mostly increases throughout development, reaching 76% expression at the onset of twitching (Figure 2f). The *in situ* data from Wormbase (Figure 1d) shows the same trend of an initial drop in TF expression as seen for *Ciona* (albeit with much lower levels of TF expression), but the *in situ* data from Wormbase may be less reliable than the *in situ* data from the databases for other organisms (many genes in Wormbase were annotated as being expressed in very broad time-ranges, such as 'embryo ' or 'adult' - genes annotated in this way were ignored for the purposes of this analysis, resulting only in subset of TFs being considered).

Evidence for TF utilization according to developmental strategy

In all six species for which *in situ* and/or microarray data was analyzed, C2H2 zinc finger proteins are over-represented during the first stage monitored, when most or all transcripts are maternally deposited. In *Xenopus tropicalis*, *Danio rerio*, *Ciona intestinalis*, *Drosophila melanogaster*, and *Anopheles gambiae*, Homeobox transcription factors are under-represented among maternal transcripts, but Homeobox TF expression increases throughout embryogenesis. The same pattern is seen in the microarray data for *Caenorhabditis elegans*, but the *in situ* data for *Caenorhabditis elegans* shows Homeobox transcription factors being slightly over-represented at the 4-cell stage. As discussed previously, however, the *Caenorhabditis elegans in situ* data may not be very reliable. Patterns of utilization for the other TF families analyzed are much less consistent. For all species other than *C. elegans* and *Ciona intestinalis*, HLH transcription factors are also under-represented among maternal and early transcripts (with the percentage of HLH transcription factors being expressed between 10 and 40% less than the overall percentage of TFs expressed) and then increase throughout development. In *C. elegans*, both the microarray and *in situ* data show HLH TFs being slightly over-represented at the 4-cell stage (16 and 31% more expressed than TFs in general). In *Ciona intestinalis*, HLH TFs are under-represented in the egg but over-represented in the 16-cell stage (26% more expressed than TFs in general), while in the other species in which HLH TFs are under-represented among the maternal transcripts, they are also under-represented during the early stage of zygotic expression.

The consistency of the pattern of Homeobox TF expression in the diverse group of species, with overall percentages of Homeobox TFs varying between species but the trend of under-representation in early stages and over-represenation in later stages holding true for all species analysed, indicates that broad patterns of Homeobox TF expression likely do not play a large role in the different modes of cell fate specification between the species. Cell fate specification in both *Caenorhabditis elegans* and *Ciona intestinalis* occurs before a large fraction of Homeobox TFs are expressed. However, HLH TFs may be important for these early developmental decisions, as HLH are over-represented among maternal transcripts or early zygotic transcripts in *Caenorhabditis elegans* and *Ciona intestinalis*, respectively. Previous work has shown that HLH TFs are involved in cell specification [33], and this study suggests that differences in the expression patterns of HLH TFs may determine the timing of cell specification.

Conserved trends of transcription factor expression

Having established an important difference in TF family utilization between species with different modes of cell fate specification, we were interested in the dominating expression trends within and between species. This is similar to work on *Drosophila* published by Hooper et al. [13], but in our case took the different developmental time scales into account.

Quantitative microarray or RNA-seq expression values for transcription factors in *Danio rerio*, *Xenopus tropicalis, Drosophila melanogaster*, *Anopheles gambiae* and *Caenorhabditis elegans* were clustered based on their temporal patterns of expression. Transcription factors which were never expressed during the time course or whose expression remained close to constant throughout all time points were filtered out prior to clustering. The remaining expression values at time-points through embryogenesis were Z-normalized so that the mean expression value was 0 and standard deviation was 1. The expression patterns were then clustered into 7 clusters using the R package *Mfuzz*, which performs fuzzy c-means clustering [34,35]. The number of clusters was empirically chosen, and adding more clusters resulted in either new clusters with very few members or clusters that were not very different from each other. Using fewer than 7 clusters generally led to the merger of previously separated clusters.

All species showed more than one cluster of TFs that was expressed/contributed maternally, with different clusters generally differing based on when expression started to decline and the rate of that decline (Figures 3, 4, and 5). All species showed several clusters with low/no initial expression that increases and then remains steady over the time course as well as one or more clusters with low/no initial expression that rises to reach a peak in expression and then decreases, usually gradually. These major types of clusters are similar to two of the three major classes of embryonic gene expression patterns noted by Hooper *et al.* [13] for *Drosophila* embryogenesis: class I (maternal) and class III (activated). The third major class observed by Hooper *et al.*, transiently expressed genes, appears to be less well-represented in our analysis. While some clusters do appear to show an increase in expression followed by a sizable decrease, not all species show such a pattern. Perhaps only a few TFs show such expression patterns or TFs with transient expression pattern generally show very individually specific periods of expression and thus do not cluster together in this analysis. It is worth mentioning that the expression classes found by Hooper et al. were based on the analysis of the entire transcriptome, and that many TF genes did not fall into any of the major three classes.

The clustered TF groups enable an alternative look at TF family utilization. In the (b) panels of Figures 3, 4 and 5, we show the degree of over- or under-representation of the Homeobox, C2H2 zinc finger, C4 zinc finger, bZIP, and HLH families, in comparison to equally sized random selections of transcription factors. The distribution of TF families among clusters supports the conclusion that C2H2 zinc fingers are important early in development and Homeobox TFs later in development, however, the resolution of this analysis is limited and does not pick up the over-representation of HLH TFs.

In order to assess the biological relevance of the expression clusters, the enrichment of GO terms in each cluster was determined using the weight01 algorithm in the R package *topGO* [36,37] (Tables S1-5). For *Drosophila melanogaster*, *Anopheles gambiae*, and *Caenorhabditis elegans*, very few GO term enrichments are significant due to the lower number of TFs in these species (and therefore, per cluster). However, GO enrichment in zebrafish and *Xenopus tropicalis* confirms the validity of our clusters. For example, in zebrafish, "cell migration involved in gastrulation" is enriched in cluster 1, which shows high initial expression that decreases gradually after the onset of gastrulation. These GO term enrichments suggest that the clusters do represent biologically relevant groupings of transcription factor genes.

We were next wondering how comparable these expression clusters were between related species within the vertebrate and invertebrate groups. In a 'clustering of clusters', the TF groups for those pairs of species were sorted using hierarchical clustering. As can be seen in the dendrograms in Figures 4a and 5a, pairs of clusters with one cluster from each of the two species emerge from such an approach. This confirms our assessment that TF expression patterns in diverged species are largely similar. For the vertebrates, all clusters form 1-to-1 pairs with a cluster in the other species; for the insects, there were only 6 pairs of clusters. For both pairs of species, the distribution of Homeobox, C2H2 zinc finger, C4 zinc finger, bZIP, and HLH TFs among the clusters was not significantly different between the two species (Fisher's Exact Test, p>0.05). The over-representation of each TF family was thus determined for each cluster pair using both species; several cluster pairs showed statistically significant over-representation of particular TF families in accordance with previously described trends.

Expression of orthologs

Having established a 'core set' of expression clusters that are represented in related species, we wanted to ascertain whether orthologs of TFs in a particular cluster in one species fall into the analogous cluster in the other species. For both pairs of species, most orthologs did not fall into analogous clusters (only 91 out of 343 pairs for the vertebrates, and 58 out of 228 for the insects). However, more orthologs fell into analagous clusters than would be expected by chance for random pairs of TFs from the two species (Z=81.35, empirical p-value $< 10^{-3}$ for the vertebrates; Z=7.44, empirical p-value $< 10^{-3}$ for the insects). To further examine the conservation of expression of TFs, the correlation between the orthologs in the two species groups at comparable time points was determined. For the vertebrates, only 193 of 343 orthologous TFs were correlated at a p-value of 0.05. Similarly, only 70 of 228 orthologous TFs for the insects were significantly correlated. This degree of correlation was higher than would be expected by chance, as determined by finding the level of correlation for random pairs of TFs from the two species (Z=10.62, empirical p-value $< 10^{-3}$ for vertebrates; Z=5.36, empirical p-value $< 10^{-3}$ for insects).

Although the level of expression conservation between TFs in each of the two pairs of species is significantly greater than would be expected from random pairs of TFs, the level of conservation for either pair of species is not very high. This lack of conservation in expression of particular TFs stands in stark contrast with the high degree of apparent conservation of major TF expression patterns. While there are key times in animal development when large changes in TF expression occur, the exact set of TFs that change in expression and play a role in development at a particular time varies between divergent species. This divergence in TF expression during development between divergent species contrasts with the similarity in TF expression during embryonic development in the *Drosophila melanogaster* subgroup. Rifkin *et al*. (2003) compared the expression of genes between *Drosophila* yakuba, *Drosophila* simulans, and 4 strains of *Drosophila melanogaster* and found that the expression of most TFs was stable across the clade, while 27% of all genes differed in their developmental expression. They suggested that changes in cis-regulatory regions, rather than changes in TFs or TF expression, drive many of the differences in developmental expression between the species [38]. However, the findings from this study indicate that for species with greater evolutionary separation, differences in TF expression may indeed play an important role in phenotypic differences during development.

The Hourglass model revisited

We extended our work to the overall similarity in the expression of all orthologous TFs at different time points during embryonic development (Figure 6). The heatmap shows the similarity in TF transcriptomes between time points in zebrafish and *Xenopus* (Figure 6a) and *Drosophila* and *Anopheles* (Figure 6b). In case of the vertebrate species, it shows that TF expression is most similar between the two species during mid/late embryogenesis, with the similarity peaking at 17 hours into development (mid/late segmentation stage) in zebrafish and *Xenopus* stage 28 (late in the early tailbud phase). Those times during development are marginally earlier than the phylotopic stage for these two species [17].

For *Drosophila melanogaster* and *Anopheles gambiae*, the time period between 16 and 18 hours in *Drosophila* and the 40 hour time point in *Anopheles gambiae* are most similar in TF expression (Figure 6). This time period is significantly later than the arthropod phylotypic stage, which is considered to be 8-10 hours into *Drosophila melanogaster* embryogenesis [28]. It can be speculated that this later time point is insect-specific, as it also correlates well with the decline in TF numbers seen in both insect species. However, there is also a local minimum in the distance between the TF transcriptomes for the two species at the 8-10 hour mark in *Drosophila*, and the first and last few hours of embryonic development show the greatest divergence in TF expression between the species. To better understand whether the similarity in TF transcriptome at the 16-18 hour time period in *Drosophila* was applicable to the transcriptome in general, the similarity of expression of non-TF orthologs was also compared for all time points in *Drosophila melanogaster* and *Anopheles gambiae*. For non-TFs, the Arthropod phylotypic period was the time period for which the transcriptomes of the two species were most similar, although the 16-18 hour time period was a local minimum for the distance between the transcriptomes.

TF expression thus mostly fits with the 'hourglass' model of development. TF expression is most different for both pairs of species early on in development and becomes more similar as embryonic development proceeds. The broad range of time, during which TF expression is fairly similar between *Drosophila* and *Anopheles* and the peak in similarity late in expression for the two species, is interesting in light of the finding by Levin *et al*. (2012) that genes up-regulated during the proposed phylotypic stage for nematodes were up-regulated during two stages in *Drosophila*, at about 6-8 hours into development (right before the Arthropod phylotypic stage) and 16-18 hours into development [39]. That second time period is the time period TF expression is most similar between *Drosophila* and *Anopheles*, and the time period during which non-TF expression is almost as similar as during the arthropod phylotypic stage. This new finding provides support for the speculation by Levin *et al*. (2012) that either the nematode phylotypic stage was sub-functionalized into two stages in insects or that two ancestral stages became coupled in nematodes. Kalinka *et al*. (2010) only used data for the first 16 hours of *Drosophila* development when comparing the transcriptomes of different *Drosophila* species and thus would have missed a second period of transcriptome similarity 16-18 hours into development.

CONCLUSIONS

The pattern of TF expression and TF family utilization throughout development appears to be largely similar in a diverse group of species including two vertebrates (*Xenopus tropicalis* and *Danio rerio*) and four invertebrates (*Ciona intestinalis*, *Caenorhabditis elegans*, *Anopheles gambiae* and *Drosophila melanogaster*). The trend of high utilization of C2H2 zinc finger TFs and low utilization of Homeobox TFs early in development with Homeobox TF utilization increasing later in development described by Adryan and Teichmann (2010) for *Drosophila melanogaster* can also be observed in the five other species considered in this study. Interestingly, there was no notable difference in the onset of Homeobox TF expression between species with a fixed cell lineages and organisms with conditional specification. However, HLH TFs, which play a role in cell specification, appear to be expressed earlier in the two species in which cell fate specification occurs early than in the species in which specification occurs later in embryonic development.

Clustering of TFs shows that within species there are sizable groups of TFs that share very similar patterns of expression. Direct comparisons of clusters between *Xenopus tropicalis* and *Danio rerio* and between *Anopheles gambiae* and *Drosophila melanogaster* show that these major patterns of TF expression are well conserved between species that have diverged 100s of millions of years ago. Despite this apparent conservation of major patterns of TF expression, expression patterns of individual TFs do not seem to be conserved. This discrepancy in conservation may suggest that the same TFs are utilized for different purposes in development in different species. For *Xenopus* and zebrafish, the expression of all TFs was most similar at approximately the vertebrate phylotypic stage, indicating that TF expression follows the 'hourglass model' in these species. For the two insect species, TF expression was most similar at a time point much later than the traditional Arthropod phylotypic stage, although TF similarity was almost as high during the phylotypic stage. This finding support the idea put forward by Levin *et al.* (2012) that the nematode phylotypic stage is split into two stages in *Drosophila*.

METHODS

Lists of transcription factor for each species

Lists of transcription factors for each species were downloaded from the DBD Transcription Factor prediction database (http://dbd.mrc-lmb.cam.ac.uk/DBD/index.cgi?Download) [32]. The files used were xn.tf.ass for *Xenopus tropicalis* (downloaded 25 June 2012), da.tf.ass for *Danio rerio* (downloaded 22 June 2012), dm.tf.ass for *Drosophila melanogaster* (downloaded 4 August 2012), cl.tf.ass for *Caenorhabditis elegans* (downloaded 15 May 2012), is.tf.ass for *Ciona intestinalis* (downloaded May 14 2012), and ag.tf.ass for *Anopheles gambiae* (downloaded 25 June 2012).

Analysis of *in situ* data

*In situ* hybridization data was obtained from the Ghost database (http://ghost.zool.kyoto-u.ac.jp/datas/GhostInsituText_201007.zip, from 1 July 2011) for *Ciona intestinalis* [40], from BDGP Release 3 (http://insitu.fruitfly.org/insitu-mysql-dump/insitu_annot.csv.gz, downloaded 1 June 2012) for *Drosophila melanogaster* [3,4], from ZFIN (http://zfin.org/downloads/wildtype-expression.txt, downloaded 28 May 2012) for *Danio rerio* [41], and from WormMart (http://caprica.caltech.edu:9002/biomart/martview/, downloaded 29 May 2012) for *Caenorhabditis elegans* [42].

For the *Ciona intestinalis* data from Ghost, clone ids had to be mapped to Ensembl protein IDs for transcription factors. A custom BLAST database was created with the cDNA sequences corresponding to the transcription factor proteins. The sequences for the clones were downloaded from the Ghost website and Blastn was used to map those sequences to the *Ciona* transcription factors. For the other species, conversions between the given identifier in the original database and Ensembl protein IDs was conducted using the *biomaRt* package in R [43].

Presence/Absence analysis of microarray and RNA-seq data

Microarray data was used for *Xenopus tropicalis*, *Danio rerio*, *Caenorhabditis elegans*, *Drosophila melanogaster*, and *Anopheles gambiae* [3,5,6,18,39]. Microarray data for most of the species was downloaded from the GEO database [9,10]. The GEO accession codes from which the data were obtained are GSE27227 for *Xenopus tropicalis*, GSE24616 for *Danio rerio*, GSE14993 for *Anopheles gambiae*, and GSE31422 for *Caenorhabditis elegans*. Microarray data for *Drosophila melanogaster* was obtained from the BDGP website (ftp://ftp.fruitfly.org/pub/embryo_tc_array_data/).

For Agilent arrays, the "gIsWellAboveBackground" flag was used to call genes as present or absent. A present call required all probes for the gene for all replicates of the condition to be called as well above background. For Affymetrix arrays, the mas5calls function in the R package *affy* [44] was used to determine presence or absence of probe sets. This function implements the MAS5 algorithm for Present/Absent calling and achieves very similar (albeit not identical) detection calls to the MAS5 software [45,46]. For a gene to be called present, every probe set for that gene for every replicate of a condition had to be called present by the mas5calls function for the gene to be called as present in that condition.

FPKM values for the modENCODE RNA-seq developmental time-course in *Drosophila melanogaster* were obtained from Supplementary Table 9 from the Graveley *et al*. (2011) publication (2010-09-11375C-TableS9revised.xls). Genes with FPKM > 1 were considered as present, the same threshold used by Graveley *et al* [19].

Pre-processing of microarray and RNA-seq data for clustering

Affymetrix microarray data was pre-processed using the R package *gcrma* [47,48]. The gcrma function was used to background correct, normalize, and summarize the raw data. For Agilent microarray data, the gProccessed signal from the Agilent Feature Extraction software output was used as the background-corrected value for each probe; the processed signal was used as Stafford and Brun (2007) found that the values obtained from the processed signal are most similar to *gcrma* background-corrected Affymetrix data[49]. Between-array quantile normalization was conducted using the limma package in R [50]. Expression values for probes matching the same gene were averaged to obtain a single expression value per gene.

For the modENCODE RNA-seq data, the $\log_2$ of the FPKM values were used as expression values for the purposes of clustering.

Clustering of microarray and RNA-seq Data

Prior to clustering, TFs that were never expressed throughout the time course (as assessed by the present/absent calling described above) were filtered out. TFs with a maximum expression less than 1 greater than the minimum expression on a $\log_2$ scale (i.e. less than two times greater) were also filtered out as being TFs with "constant expression". All remaining TFs were Z-normalized using the standardise function in *Mfuzz* [34,35]. Clustering was then performed with the *mfuzz* function with the number of clusters set to 7 and the fuzzifier coefficient M set to 1.25. Genes with a membership score greater than 0.8 were considered to be part of a cluster.

Statistical significance of over- or under- representation of members of a TF family in a cluster was determined using a hypergeometric test. The p-value was set at 0.0025 for both one-sided tests. This low p-value was selected to take into account that 10 tests were being performed for each TF family.

GO terms associated with TFs for each species were obtained from Ensembl via the *biomaRt* package in R. Enrichment of terms within clusters was determined using the *topGO* package in R [36,37]. The weight01 algorithm was used. A p-value of 0.01 was used to assess significance.

Comparisons of clusters in different species

Comparing clusters between *Xenopus tropicalis* and *Danio rerio* and between *Anopheles gambiae* and *Drosophila melanogaster* required aligning the time points in the time courses for the two species being compared. For *Xenopus tropicalis* and *Danio rerio*, the alignment of time points is described in Supplementary Figure S3.

For *Drosophila melanogaster*, the time course consisted of 12 two-hour time intervals between 0 and 24 hours. The 12 time points that were selected as corresponding to those 12 points in *Anopheles gambiae* were the 2, 6, 10, 16, 19, 22, 28.5, 31, 34, 40, 43, and 49 hour time points. These time points are not perfectly spaced out, which would have been more ideal, but they correspond to approximately the same time interval in *Drosophila melanogaster* assuming a linear relation between developmental time in the two species.

For the clustering of the clusters, hierarchical clustering was performed using the Euclidean distances between the centers of each of the clusters. Additionally, pairing of clusters was determined by establishing whether the distance between the two clusters in a proposed pair was less than the distance between either member of the pair and any other cluster.

Comparison of ortholog expression

Lists of orthologs between species were downloaded from the InParanoid database (http://inparanoid.sbc.su.se/download/) (Ostlund *et al.* 2010). Pairs of orthologs listed as the unique seed orthologs for an ortholog were considered as orthologs for this study. Cluster membership for pairs of orthologs was compared in order to determine whether orthologs fell into clusters that were paired in the cluster comparison. A similar analysis was performed on a 1,000 sets of random pairs of genes in order to determine the significance of the findings for the orthologs.

Correlation of expression for pairs of TF orthologs was determined by using the expression values for the "common" time points between the two species. The similarity of all TF expression between time points for the different organism was determined by finding the Euclidean distance between the vectors of Z-normalized ortholog expression for all time points.

Acknowledgments:

We thank Marc Jones for contributing to a pilot study that initiated this work.

# Supplemental Tables

Table 1: GO Terms significantly enriched in zebrafish clusters at $p < 0.01$

| Cluster | GO.ID | Term | Annotated | Significant | weight |
|---:|---|---|---:|---:|---:|
| 1 | GO:0042074 | cell migration involved in gastrulation | 11 | 6 | 0.0029 |
| 2 | GO:0007275 | multicellular organismal development | 323 | 45 | 0.0013 |
| 3 | GO:0060788 | ectodermal placode formation | 13 | 6 | 0.0012 |
| 3 | GO:0035121 | tail morphogenesis | 3 | 3 | 0.0013 |
| 5 | GO:0007275 | multicellular organismal development | 323 | 94 | 2.6e-14 |
| 5 | GO:0043010 | camera-type eye development | 38 | 16 | 0.0052 |
| 5 | GO:0006355 | regulation of transcription, DNA-depende... | 612 | 119 | 0.0066 |
| 7 | GO:0007165 | signal transduction | 133 | 27 | 0.0088 |
| 7 | GO:0043401 | steroid hormone mediated signaling pathw... | 56 | 14 | 0.0089 |
| 9 | GO:0043524 | negative regulation of neuron apoptosis | 2 | 1 | 0.0086 |

Table 2: GO Terms significantly enriched in *Xenopus tropicalis* clusters at $p < 0.01$

| Cluster | GO.ID | Term | Annotated | Significant | weight |
|---:|---|---|---:|---:|---:|
| 1 | GO:0016573 | histone acetylation | 4 | 3 | 0.0059 |
| 2 | GO:0007067 | mitosis | 2 | 2 | 0.0071 |
| 2 | GO:0051301 | cell division | 4 | 2 | 0.0381 |
| 4 | GO:0045944 | positive regulation of transcription fro... | 20 | 10 | 1.1e-05 |
| 4 | GO:0009952 | anterior/posterior pattern specification | 22 | 14 | 0.00035 |
| 4 | GO:0014034 | neural crest cell fate commitment | 8 | 5 | 0.00061 |
| 4 | GO:0045665 | negative regulation of neuron differenti... | 3 | 3 | 0.00131 |
| 4 | GO:0030902 | hindbrain development | 11 | 8 | 0.00143 |
| 4 | GO:0001756 | somitogenesis | 7 | 5 | 0.00166 |
| 4 | GO:0030917 | midbrain-hindbrain boundary development | 6 | 4 | 0.00175 |
| 4 | GO:0045666 | positive regulation of neuron differenti... | 6 | 4 | 0.00175 |
| 4 | GO:0051090 | regulation of sequence-specific DNA bind... | 14 | 6 | 0.00376 |
| 4 | GO:0042472 | inner ear morphogenesis | 13 | 7 | 0.00420 |
| 4 | GO:0033504 | floor plate development | 6 | 4 | 0.00465 |
| 4 | GO:0006916 | anti-apoptosis | 4 | 3 | 0.00481 |
| 4 | GO:0000122 | negative regulation of transcription fro... | 17 | 6 | 0.00686 |
| 4 | GO:0043049 | otic placode formation | 8 | 4 | 0.00686 |
| 5 | GO:0007275 | multicellular organismal development | 192 | 56 | 3.5e-11 |

Table 3: GO Terms significantly enriched in worm clusters at $p < 0.01$

| Cluster | GO.ID | Term | Annotated | Significant | weight |
|---:|---|---|---:|---:|---:|
| 9 | GO:0016048 | detection of temperature stimulus | 1 | 1 | 0.0059 |

Table 4: GO Terms significantly enriched in mosquito clusters at $p < 0.01$

| Cluster | GO.ID | Term | Annotated | Significant | weight |
|---:|---|---|---:|---:|---:|
| 1 | GO:0006978 | DNA damage response, signal transduction... | 2 | 2 | 0.0042 |
| 1 | GO:0010165 | response to X-ray | 2 | 2 | 0.0042 |
| 1 | GO:0010332 | response to gamma radiation | 2 | 2 | 0.0042 |
| 1 | GO:0031571 | mitotic cell cycle G1/S transition DNA d... | 2 | 2 | 0.0042 |
| 1 | GO:0034644 | cellular response to UV | 2 | 2 | 0.0042 |
| 1 | GO:0042771 | DNA damage response, signal transduction... | 2 | 2 | 0.0042 |
| 1 | GO:0043523 | regulation of neuron apoptosis | 2 | 2 | 0.0042 |
| 6 | GO:0007179 | transforming growth factor beta receptor... | 3 | 3 | 0.00083 |
| 8 | GO:0045944 | positive regulation of transcription fro... | 5 | 4 | 0.0068 |

Table 5: GO Terms significantly enriched in fly clusters at $p < 0.01$

| Cluster | GO.ID | Term | Annotated | Significant | weight |
|---|---|---|---|---|---|
| 2 | GO:0000022 | mitotic spindle elongation | 7 | 3 | 0.0018 |
| 2 | GO:0007186 | G-protein coupled receptor signaling pat... | 25 | 6 | 0.0034 |
| 2 | GO:0045944 | positive regulation of transcription fro... | 3 | 2 | 0.0046 |
| 2 | GO:0050911 | detection of chemical stimulus involved ... | 4 | 2 | 0.0090 |
| 6 | GO:0007591 | molting cycle, chitin-based cuticle | 2 | 2 | 0.0036 |
| 6 | GO:0007367 | segment polarity determination | 3 | 2 | 0.0104 |
| 8 | GO:0046716 | muscle cell homeostasis | 3 | 2 | 0.0057 |

Figure legends

Figure 1. TF and TF family expression through embryogenesis based on *in situ* hybridization data. The gray vertical line in each plot indicates the onset of gastrulation.

Figure 2. TF and TF family expression throughout embryogenesis based on microarray or RNA-seq data. The gray vertical line in each plot indicates the onset of gastrulation.

Figure 3. Clustering of transcription factor expression patterns in *C. elegans*. (a) The expression patterns of members of each of the seven clusters determined by Mfuzz. The gray vertical line indicates the onset of gastrulation. (b) The relative representation of different TF families in each cluster. Blue stars signify statistically significant over-representation of a TF family in a cluster; red stars signify statistically significant under-representation of a TF family in a cluster.

Figure 4. Comparison of TF expression clusters in *Xenopus tropicalis* (XT) and *Danio rerio* (DR).(a) The expression profiles of the cluster pairs. The gray vertical line indicates the onset of gastrulation. The plot in the third column shows the values of the centers of the clusters at each of the common time points. Orange is *Xenopus tropicalis* and blue is *Danio rerio*. (b) Dendrogram showing hierarchical clustering of the clusters. (c) TF family representation in each cluster. The dark gray represents *Xenopus tropicalis* and the light gray *Danio rerio*. Blue stars signify statistically significant over-representation of a TF family in a cluster pair; red stars signify statistically significant under-representation of a TF family in a cluster pair.

Figure 5. Comparison of TF expression clusters for *Drosophila melanogaster* and *Anopheles gambiae*. (a) The expression profiles of the cluster pairs. The gray vertical line indicates the onset of gastrulation. The plot in the third column shows the values of the centers of the clusters at each of the common time points. Orange is *Drosophila melanogaster* and blue is *Anopheles gambiae*. (b) Dendrogram showing hierarchical clustering of the clusters. (c) TF family representation in each cluster. The dark gray represents *Drosophila melanogaster* and the light gray *Anopheles gambiae*. Blue stars signify statistically significant over-representation of a TF family in a cluster pair; red stars signify statistically significant under-representation of a TF family in a cluster pair. (d) Clusters that were not paired.

Figure 6. Similarity of TF transcriptomes between species. Heatmaps showing the similarity of the TF transcriptome at all time points for (a) *Xenopus tropicalis* and *Danio rerio* and (b) *Drosophila melanogaster* and *Anopheles gambiae*. Darker blue/violet shading indicates that the TF trancriptomes for the two species at the given times during development are more similar. In (b), the bracketed time periods for *Drosopila* are those included in the study by Kalinka et al. (2010) that showed that the time period between 8-10 hours is when the transcriptomes of different *Drosophila* species are most similar. That time period (indicated by a *) is a local maximum for TF transcriptome similarity between the two insects considered in this study; the 16-18 hour time period for *Drosophila melanogaster* (indicated by a **) is the time period with the greatest TF transcriptome similarity to *Anopheles gambiae*.

Figure S1: Evolutionary relationship between species considered in study.

Figure S2: Distribution of TF families for TFs in DBD database for the species under consideration in this study.

Figure S3: Time points for the zebrafish and *Xenopus tropicalis* microarray time courses. The gray dashed linesconnect time points in the two species that were considered as common time points. Spacing between stages for the *Xenopus tropicalis* timeline is based on approximate times between the onset of each stage.

Figure 1

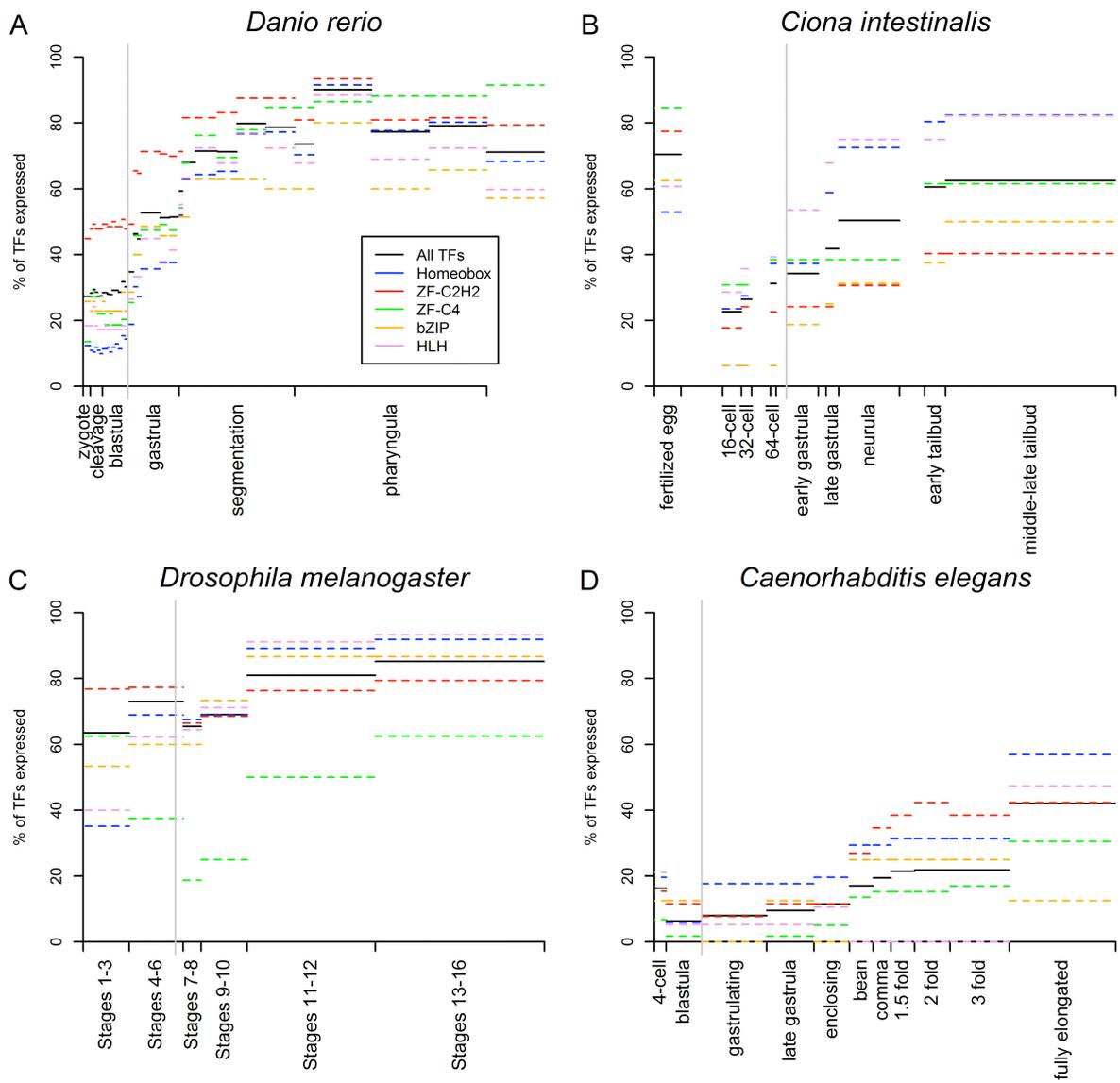

Figure 2

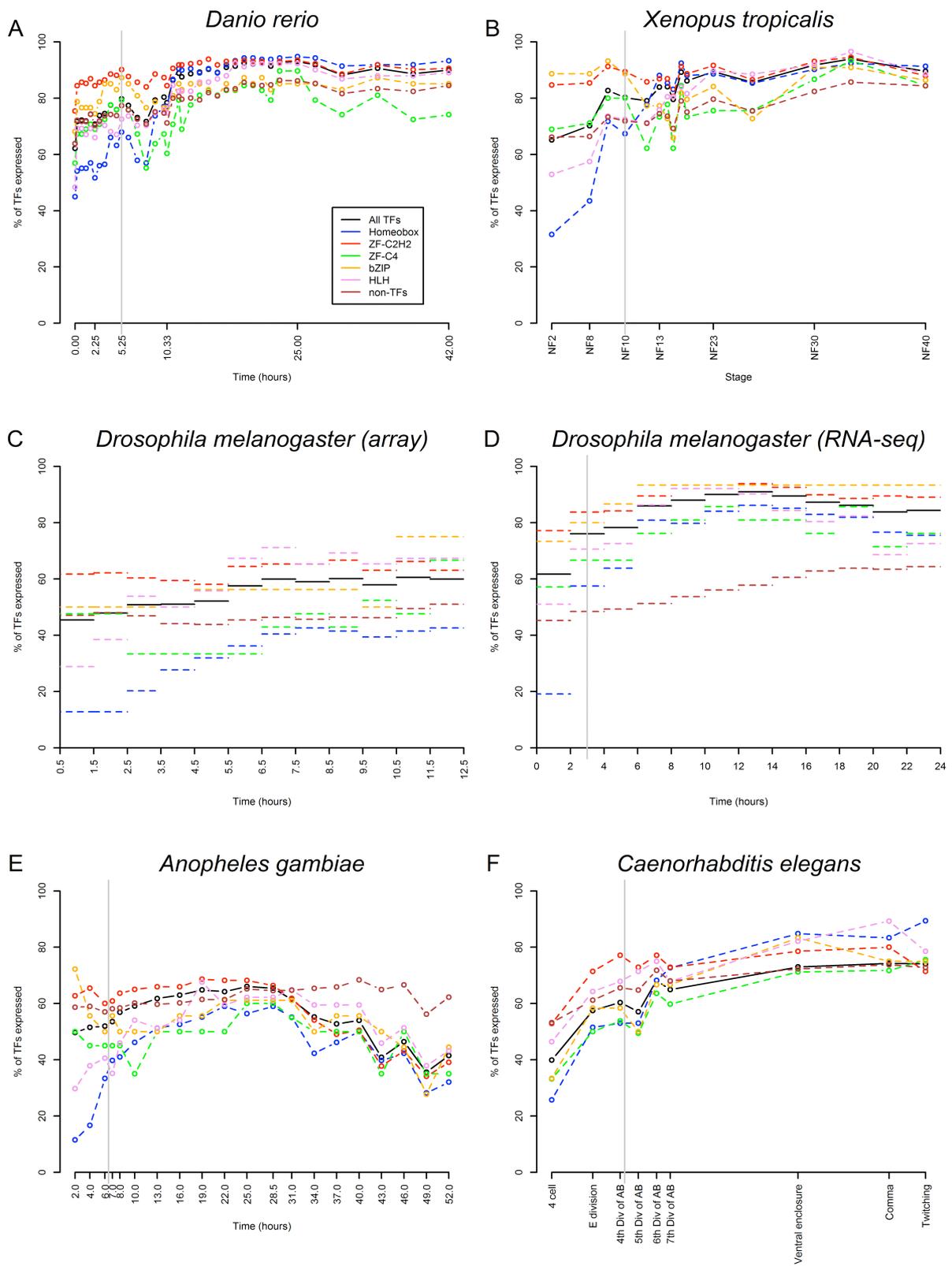

Figure 3

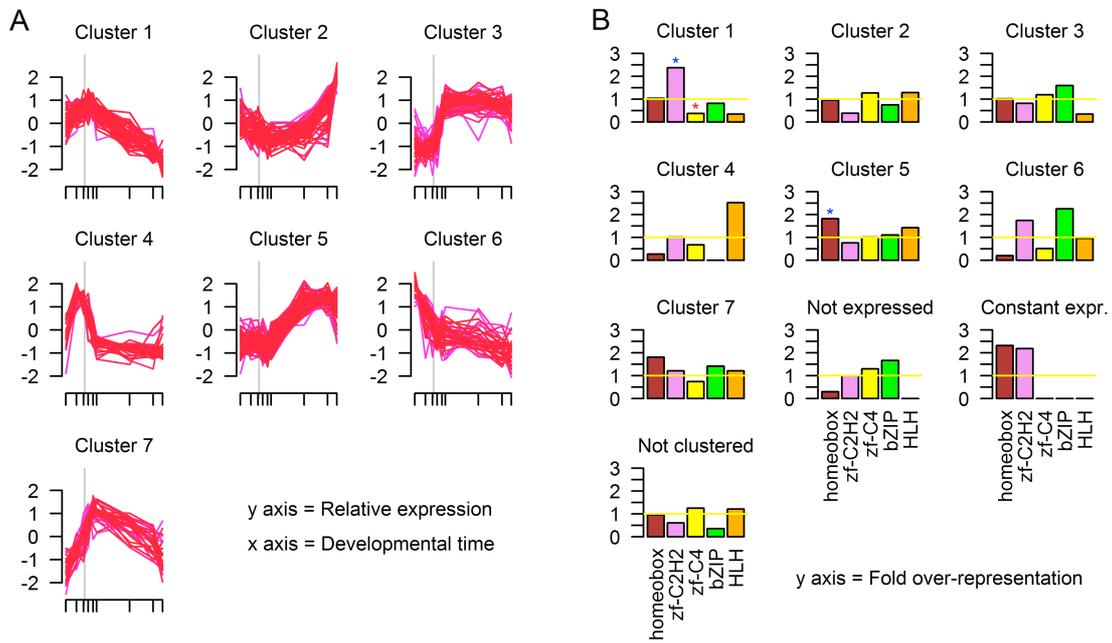

Figure 4

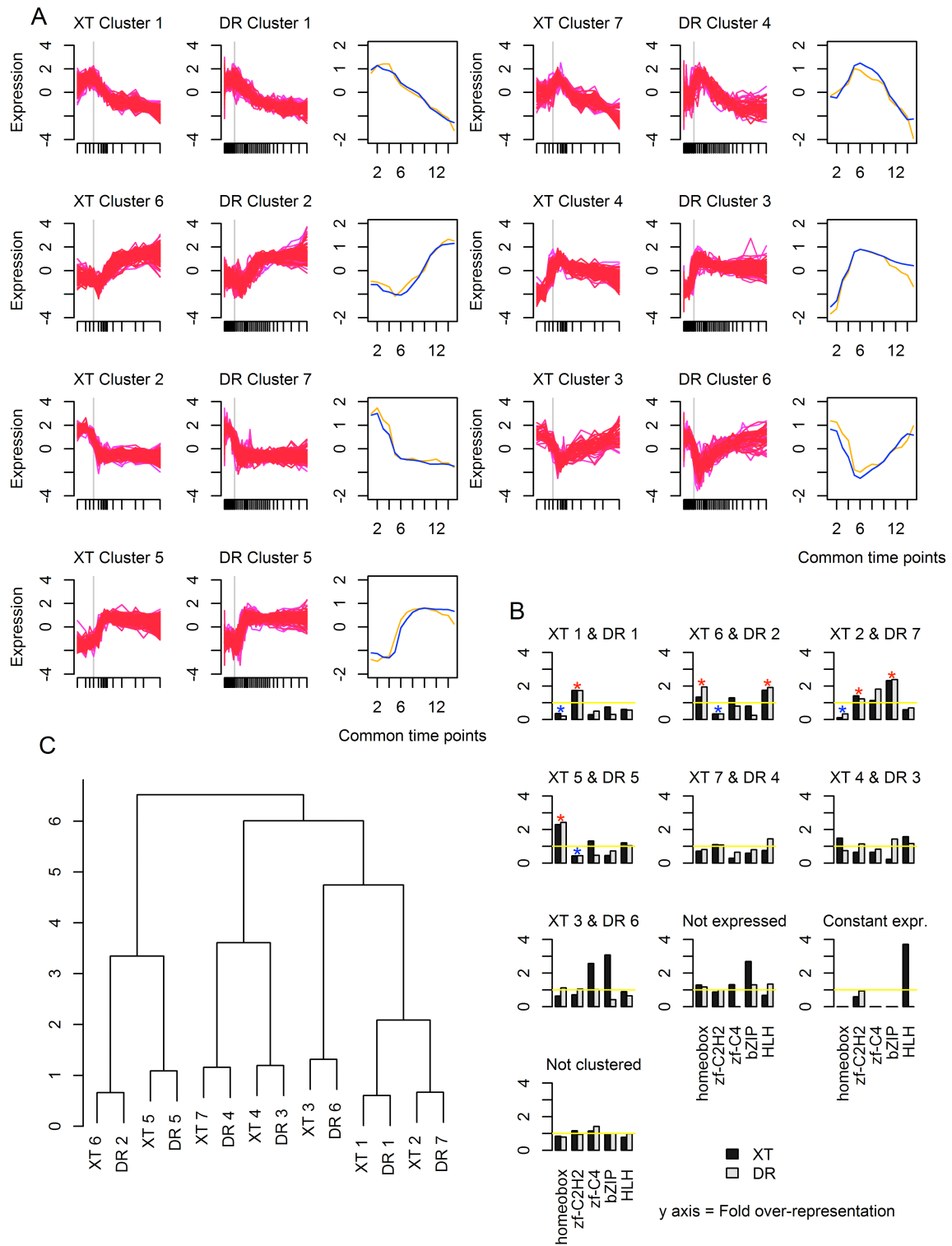

Figure 5

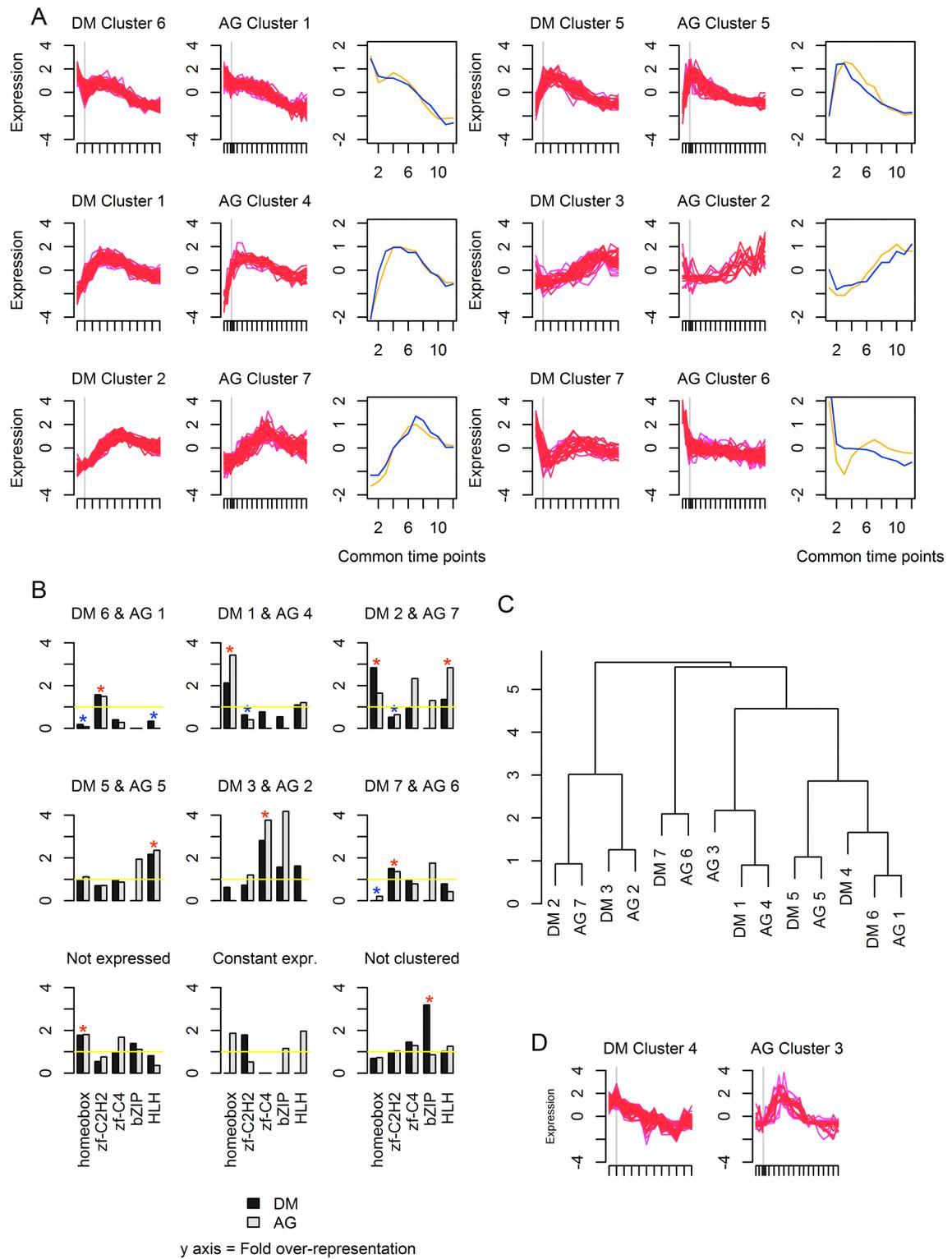

Figure 6

A

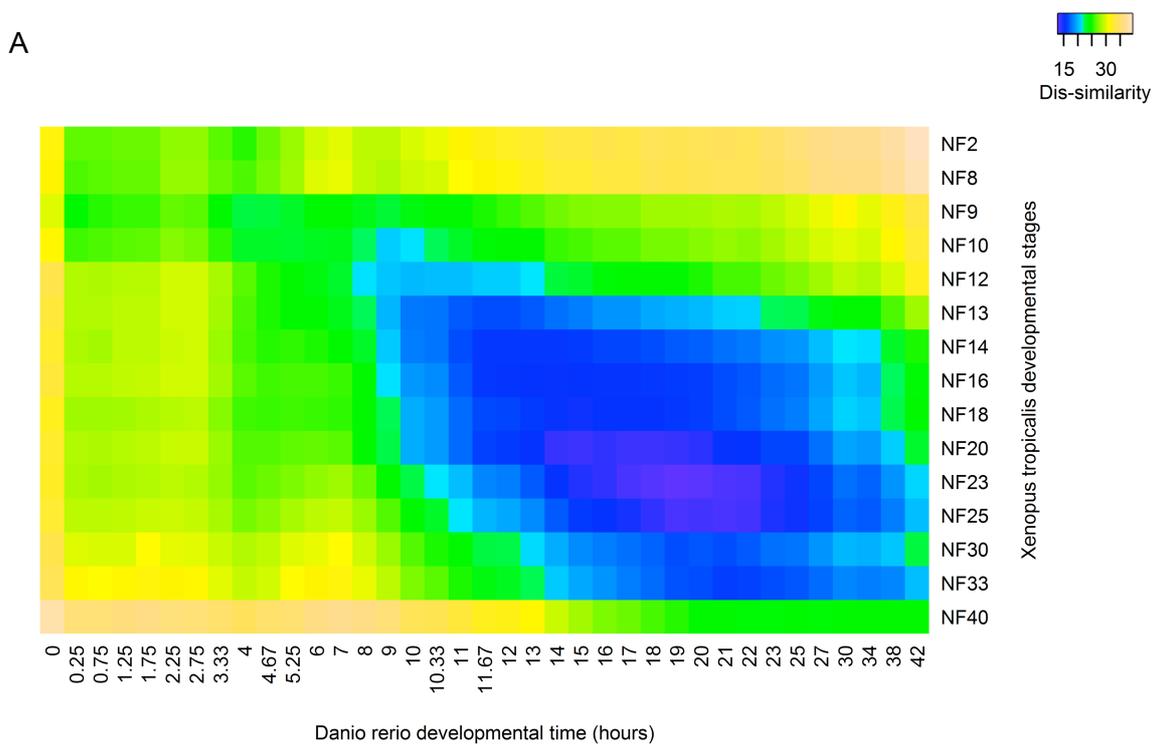

B

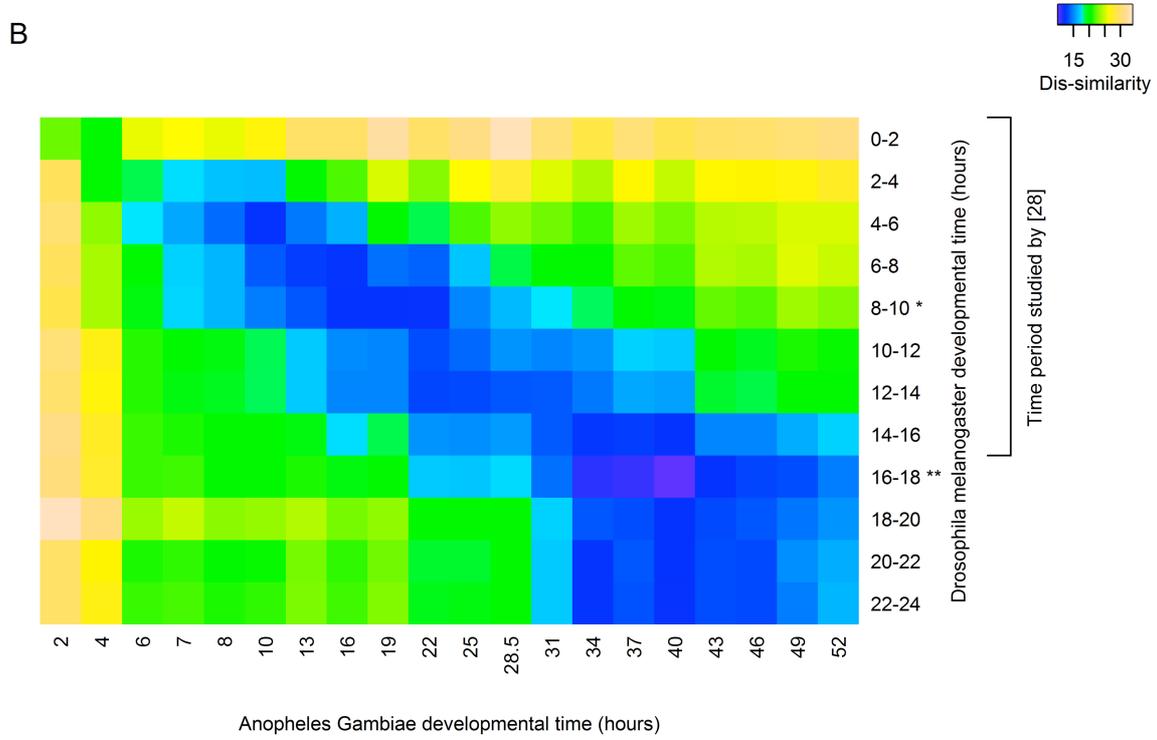

Suppl. Figure S1

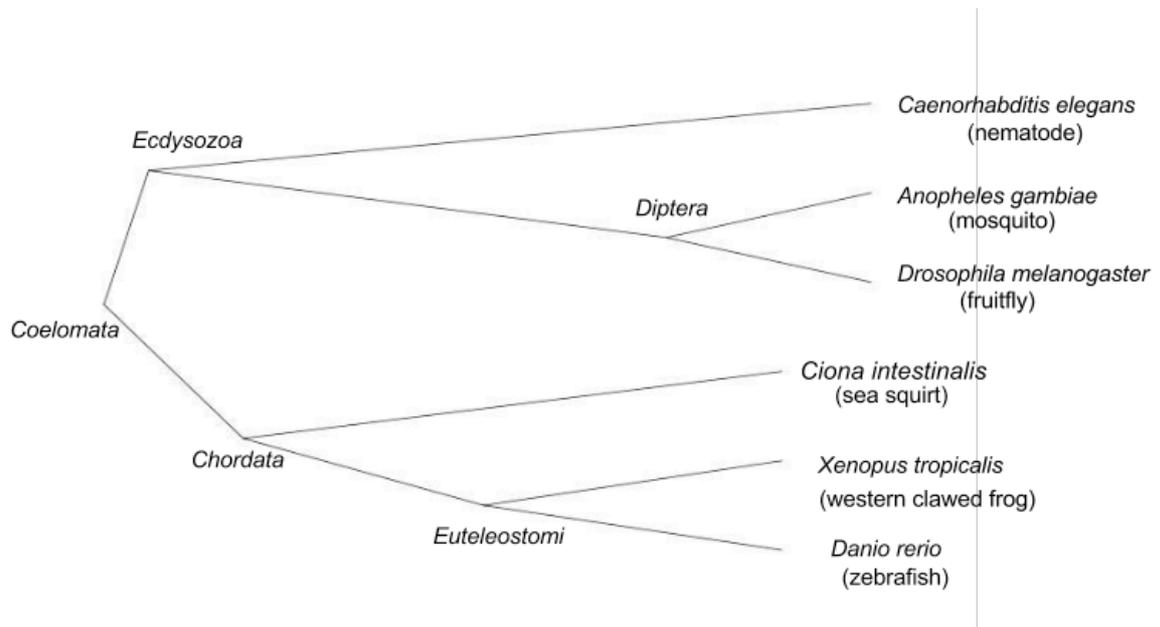

Suppl. Figure S2

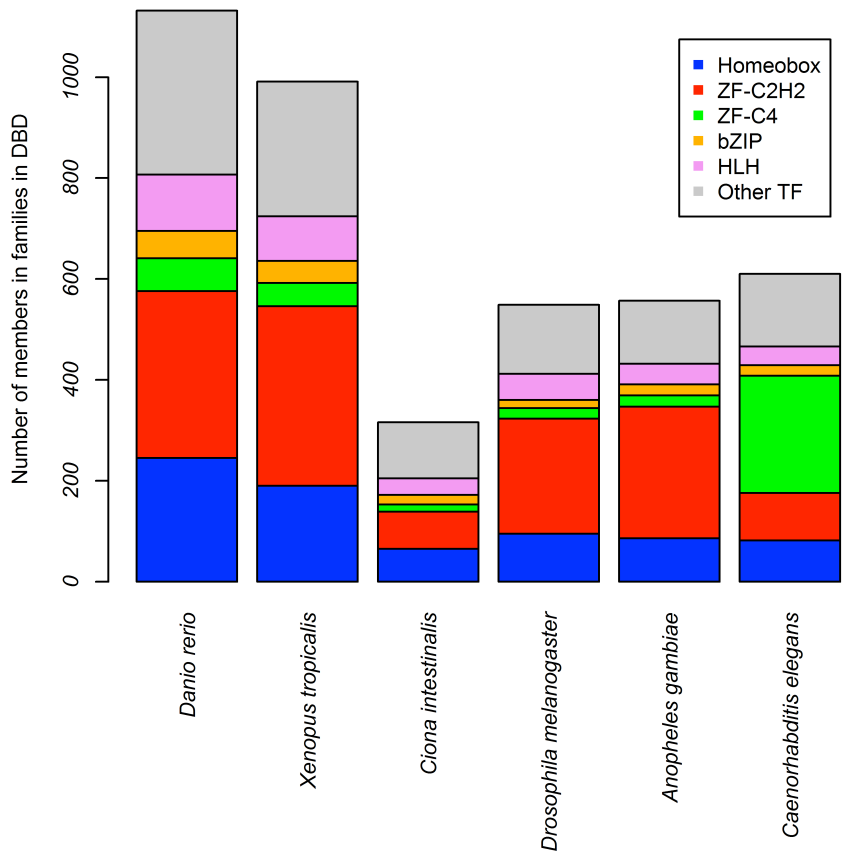

Suppl. Figure S3

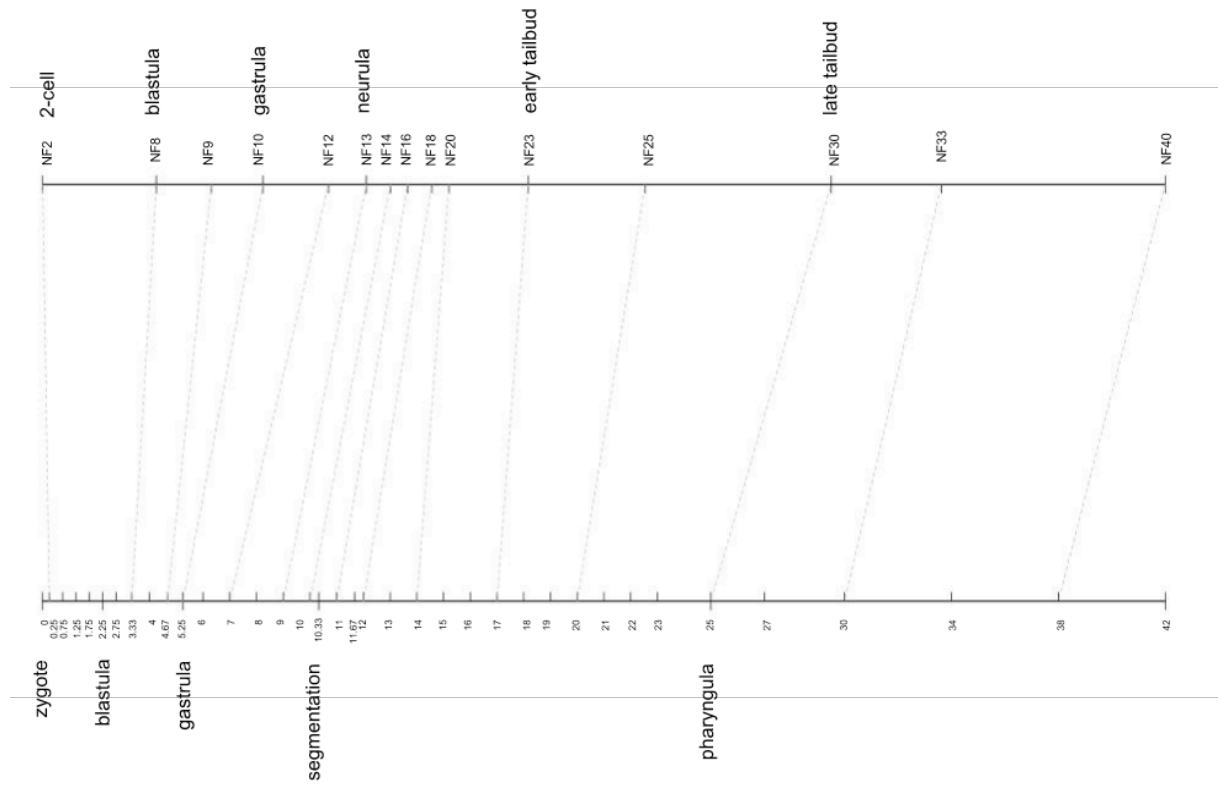